\RequirePackage{fix-cm}
\documentclass[smallextended]{svjour3}       
\smartqed  
\usepackage{graphicx}
\usepackage{ae}
\usepackage{amsmath}
\usepackage{amsfonts}
\usepackage{fontenc}
\usepackage{float}
\usepackage{hyperref}
\usepackage{tikz}

\begin{document}

\title{Mesoscopic description of the adiabatic piston: kinetic equations and $\mathcal H$-theorem
}
\author{Nagi Khalil 
}


\institute{Nagi Khalil \at
  IFISC (CSIC-UIB), Instituto de F\'isica Interdisciplinar y Sistemas Complejos, \\
  Campus Universitat de les Illes Balears, E-07122, Palma de Mallorca, Spain \\
  Tel.:  +34971172008\\
  Fax: +34971173248\\
  \email{nagi@ifisc.uib-csic.es}           
}

\date{Received: date / Accepted: date}

\maketitle

\begin{abstract}
The adiabatic piston problem is solved at the mesoscale using a Kinetic Theory approach. The problem is to determine the evolution towards equilibrium of two gases separated by a wall with only one degree of freedom (the adiabatic piston). A closed system of equations for the distribution functions of the gases conditioned to a position of the piston and the distribution function of the piston is derived from the Liouville equation, under the assumption of a generalized molecular chaos. It is shown that the resulting kinetic description has the canonical equilibrium as a steady-state solution. Moreover, the Boltzmann entropy, which includes the motion of the piston, verifies the $\mathcal H$-theorem. The results are generalized to any short-ranged repulsive potentials among particles and include the ideal gas as a limiting case.   
\end{abstract}

\clearpage

\section{Introduction}
\label{intro}

The adiabatic piston problem is an old problem in Thermodynamics and Statistical Mechanics. As described by Callen \cite{ca63} many years ago, this is to predict the equilibrium state of a cylinder containing two subsystems (usually two gases) separated by a movable adiabatic piston, i.e. a wall that isolates the two subsystems when kept fixed. The same author, and others \cite{cule79}, stated that Thermodynamics can only provide a partial solution to the problem, namely the equality of pressures. 

Lieb \cite{li99} went one step further by pointing out that a Statistical Mechanics solution to the problem is in contradiction with Thermodynamics. Namely, as described by Landau and Lifshitz \cite{lali13} and by Feynmann \cite{fe11}, the relaxation of the system, through the motion of the piston, is towards a global equilibrium, where all subsystems including the piston have the same temperature and the pressures of the gasses are equal. But this implies that the initially hotter subsystem decreases its entropy by doing work to the other subsystem, which may be in contradiction with the Second Principle \cite{glpr71,chlesi02a,grma13}. According to Lieb and Yngvason \cite{li99,liyn98,liyn99} the adiabatic piston problem can be solved by reconsidering the principle of maximizing the entropy in the absence of constrains. 

Most works on the adiabatic piston use simple models, such as non-interacting particles. In this context, rigorous results can be found in the literature. If the initial condition is of a fixed piston surrounded by two ideal gases in equilibrium with the same pressures and temperatures, the dynamics of the piston occurs in several stages \cite{chle02}: the initial condition becomes unstable after a short time and the piston start oscillating \cite{si99,nesi04,wr07}, the amplitude of the oscillations attenuates exponentially, and a final thermalization leads to the equilibration of the whole system. For more general initial conditions, where the two gases may be in different states, the evolution is in two stages \cite{chlesi02a,grpale03,cachle04}: a first fast evolution towards mechanical equilibrium, where the two gases have the same pressure, is followed by a much slower relaxation towards thermal equilibrium, dominated by the energy flux through the piston. Moreover, if the effect of recollisions (when a particle collide with the piston more than once) can be removed under some limits, the motion of the piston converges to an Ornstein-Uhlenbeck process \cite{chlesi02a,ho71,dugole81}. See \cite{chlesi02a} for an attempt to include recollisions.

The last stage of the dynamics, under the condition of mechanical equilibrium but with the temperatures of the gases still different, has attained much attention. As a matter of fact, the existence of this stage is closely related to the original controversy of the adiabatic piston problem. An interesting model, that keeps the system always in this last stage, considers two semiinfinite ideal gases in each sides of the piston having the same pressure and temperature difference. Even the fact that the net force to the piston is zero, the system reaches a steady state where the mean velocity of the piston is constant, different from zero, and towards the hotter gas \cite{grpi99,pigr99,grfr99,pi01,ch04,itsa15}. When the mass of the piston goes to infinity, its mean velocity vanishes \cite{pi02,grpa02,grpale02}. See also \cite{grpale04} for a kinetic description of the model in close relation with Thermodynamics. 

From a mesoscopic and macroscopic scales perspective, the major challenge when dealing with the adiabatic piston is to derive  closed equations for the piston and the gases. This has been done rigorously \cite{lepisi00,chlesi02} and approximately \cite{crdise96,gr99,cepapivu07,gi10,mavake05,magaba06} for the case of non-interacting and similar models. For more realistic ones, such as models with hard spheres or disks, most of the existing results are numerical \cite{hure06,whgove02}, or use rude approximations \cite{brkh12}. As in the case of non-interacting particles, the numerical simulations show an exponential relaxation of the system towards thermal equilibrium \cite{kevama00,fosh13}. 

So far, the numerous theoretical results have not been compared against experiments. This is partly because the last stage of the evolution, actually the interesting one for the adiabatic piston problem, is too much slow to be observed in macroscopic system. As far as we know, there is only one experimental work that can be relevant for the present discussion \cite{leda10}. It provides experimental measurements on the thermalization of a piston surrounded by a few hundreds of grains  in a very dense configuration, close to the jamming transition. What we learn from this experiment, and also from the theory done in the dilute regime \cite{brkh11}, is that the equilibration in the inelastic or granular case is completely different from the classic or elastic one. Specifically, when there is no energy source supplying the dissipation of the inelastic collisions, each of the gases reaches a state close to the homogeneous cooling state \cite{kh18}, and the ``thermal'' equilibration occurs when the cooling rates (or rates of energy loss) of both gases and the piston are the same.  The new ``equilibrium'' criterion, as opposite of the temperatures being the same for the elastic case, explains the emergence of non-equilibrium phase transitions \cite{brrebr05,hure06a,brkh10}. See also \cite{brkh11,brru04,brru08,brru09,brru09,brru10} for theoretical and numerical studies relevant for experimental situations. 

The works on the adiabatic piston in contact with granular matter suggest the need to include at least two new ingredients in the study of the (elastic) adiabatic piston. Firstly, if we are interested in making some contact with reality, a systematic derivation of a theory for the piston in contact with a finite number of particles is needed. Secondly, the theory should include interactions between particles. The aim of the present work is to propose a mesoscopic description of a realistic model of the adiabatic piston. The model considers a large but finite number of interacting particles, and reduces to the case of non-interacting particles when the diameters of the particles tend to zero. 

The organization of the work is as follow. The model is presented in Sec. \ref{sec:2} where we  also show, by computing some relevant quantities of the system at thermal equilibrium, the importance of keeping the spatial correlations between the gases and the piston. Section \ref{sec:3} contains the main results of the work, namely the derivation of a closed set of kinetic equations for the gases, conditioned to a position of the piston, and the piston. It is shown that the equations support the equilibrium solution of Sec. \ref{sec:2}. As an application to the results of Sec. \ref{sec:3}, in Sec. \ref{sec:4} we demonstrate the $\mathcal H$-theorem, which implies the evolution of the system towards thermal equilibrium from any initial condition. Finally, Sec. \ref{sec:5} includes a discussion together with some conclusions.

\section{Model and equilibrium \label{sec:2}}
\subsection{Model}
\begin{figure}[!h]
  \centering
  \includegraphics[width=.5\textwidth]{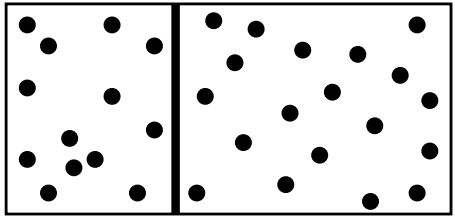}
  \caption{A representation of the system for $d=2$. }
  \label{fig:1}
\end{figure}

The system is a $d$-dimensional cylinder of length $L$ and section $S$ divided into two parts by a moving wall of section $S$, zero width, and mass $m_p$ (the piston). The normal direction of the piston remains parallel to the axis of the cylinder, taken as the $X$-axis. The system contains $N_i$ $d-$dimensional hard spheres with masses $m_i$ and diameter $\sigma_i$, to the left of the piston for $i=1$ and to the right of the piston for $i=2$. See figure \ref{fig:1} for a sketch of the system. 

The microstate of the system is given by the set of positions and velocities as $\{\mathbf r_1,\dots,\mathbf r_{N},x_p;\mathbf v_1,\dots,\mathbf v_{N},v_p\}\equiv \{\mathbf R,x_p;\mathbf V,v_p\}$ or the set of positions and momenta $\Gamma\equiv \{\mathbf r_1,\dots,\mathbf r_{N},x_p;\mathbf p_1,\dots,\mathbf p_{N},v_p\}\equiv \{\mathbf R,x_p;\mathbf P,p_p\}$, where $\mathbf p_i=m_i\mathbf v_i$, $\mathbf p_p=m_p\mathbf v_p$,  and $N=N_1+N_2$ is the total number of particles. In the latter expressions, the indexes $i=1,\dots,N_1$ stand for the particles to the left of the piston and $i=N_1+1,\dots,N$ for the particles to the right. 

The state of the system changes because of the free motion of the particles and the piston, as well as of instantaneous collisions among particles and between particles and the piston. The collisions conserve energy and linear momentum. More specifically, for two particles colliding with velocities $\mathbf v_k$ and $\mathbf v_l$, their postcollisional velocities, denoted by a prime, are
\begin{eqnarray}
  \label{eq:coll}
    && \mathbf{v}_k'=\mathbf{v}_k-[(\mathbf{v}_k-\mathbf{v}_l)\cdot \hat{\boldsymbol{\sigma}}]\hat{\boldsymbol{\sigma}}, \\
  \label{eq:coll2}
    && \mathbf{v}_l'=\mathbf{v}_l+[(\mathbf{v}_k-\mathbf{v}_l)\cdot \hat{\boldsymbol{\sigma}}]\hat{\boldsymbol{\sigma}},
\end{eqnarray}
where $\hat{\boldsymbol{\sigma}}$ is a unit vector pointing from the center of particle $k$ to the center of particle $l$ when in contact. The collision rule for a particle with mass $m_i$ and velocity $\mathbf v$ and the piston with velocity $v_p$  is 
\begin{eqnarray}
  &&v_x'= v_x-\frac{2m_p}{m_i+m_p}(v_x-v_p), \\
  &&\mathbf v_\perp'= \mathbf v_\perp, \\
  &&v_p'= v_p-\frac{2m_i}{m_i+m_p}(v_p-v_x),
\end{eqnarray}
when $v_x>v_p$ if the particle is on the left ($i=1$) and $v_x<v_p$ if the particle is on the right ($i=2$). The symbol $\perp$ represents the component of the velocity normal to the $X$-axis. An alternative characterization of the dynamics, more suitable when using the phase space representation, is provided by the hamiltonioan
\begin{equation}
  \label{eq:ham}
  \mathcal H =\left\{
    \begin{split}
      & \sum_{i=1}^{N_1}\frac{p_i^2}{2m_1}+\sum_{i=N_1+1}^{N}\frac{p_i^2}{2m_2}+\frac{p_p^2}{2m_p}, \quad (\mathbf R,x_p)\in \mathcal V, \\
      & +\infty, \qquad (\mathbf R,x_p)\notin \mathcal V,
    \end{split}
  \right.
\end{equation}
where $\mathcal V$ is the volume of the phase space accessible to the system (no overlaps).

The main focus in this work is on a mesoscopic description of the system. First, we consider the canonical equilibrium and the probability density of the phase space $\rho(\mathbf R,x_p;\mathbf P,p_p)$ and related probability densities. Later, a non-equilibrium description based on Kinetic Theory will be proposed and the distribution functions for the gases and the piston, to be defined later, will be used instead. Many of the results to be obtained at equilibrium, although of trivial derivation, will deserve as a reference and a guide for a more general out-of-equilibrium description.

\subsection{Canonical equilibrium}

If the system is big enough or if it is in contact with a thermal bath with temperature $T$, i.e. if the walls of the container vibrates in equilibrium with temperature $T$, then the probability density of the whole system is given by the canonical distribution \cite{pabe11}
\begin{equation}
  \rho=\frac{e^{-\beta \mathcal H}}{h^{dN+1}\mathcal Z},
\end{equation}
where $h$ is the Planck constant, $\beta\equiv \frac{1}{k_BT}$ with $k_B$ the Boltzmann constant, $\mathcal H$ is the hamiltonian of Eq. \eqref{eq:ham}, and $\mathcal Z$ the partition function. The latter is difficult to compute in general, due to the exclusion effects. However, for $N(\sigma_1+\sigma_2) \ll L$, that is if we neglect the volume of particles, an approximation to be taken along the work, the canonical distribution is well approximated as 
\begin{eqnarray}
  \nonumber
  \mathcal Z=&&h^{-(dN+1)} \int d\mathbf R dx_p d\mathbf P dp_p  e^{-\beta \mathcal H} \\ \simeq&& \nonumber \frac{(LS)^{N}L}{\lambda_{1}^{dN_1}\lambda_{2}^{dN_2}\lambda_{p}} \int_{0}^1 ds \left[\int_0^s dx\right]^{N_1}\left[\int_s^1 dy\right]^{N_2} \\
  =&& \frac{(LS)^{N}L}{\lambda_{1}^{dN_1}\lambda_{2}^{dN_2}\lambda_{p}} \frac{N_1!N_2!}{(N+1)!}.
\end{eqnarray}
where $\lambda_{i}\equiv \sqrt{\frac{\beta h^2}{2\pi m_i}}$ is the thermal length associated to a mass $m_i$, $i=1,2,p$.

From the probability density we can obtain other relevant quantities, such as the probability density of a particle to the left $\rho_1$, to the right $\rho_2$, and of the piston $\rho_p$. For the first one, we have 
\begin{eqnarray}
  \nonumber    \rho_1(\mathbf r,\mathbf p)\equiv &&\int \prod\limits_{i=2}^{N}d\mathbf r_id\mathbf p_i  dx_p dp_p  \rho(\mathbf R, x_p;\mathbf p,p_p)\\   \nonumber \simeq &&\frac{\lambda_{1}^de^{-\beta\frac{p^2}{2m_1}}}{h^dLS}\frac{(N+1)!}{N_1!N_2!}  \int_{x/L}^1 ds\left[\int_0^s dx\right]^{N_1-1}\left[\int_s^1 dy\right]^{N_2}\\ =&&\frac{(N+1)\lambda_{1}^d}{N_1h^dLS}\left[1-I\left(\frac{x}{L},N_1,N_2+1\right)\right]e^{-\beta\frac{p^2}{2m_1}},
\end{eqnarray}
where $I(x;a,b)$ is the regularized incomplete beta function. Proceeding analogously,
\begin{eqnarray}
  && \rho_2(\mathbf r,\mathbf p)\simeq \frac{(N+1)\lambda_{2}^d}{N_2h^dLS}I\left(\frac{x}{L},N_1+1,N_2\right)e^{-\beta\frac{p^2}{2m_2}}, \\
  \label{eq:rhop}
  && \rho_p(x,p)\simeq \frac{(N+1)!\lambda_{p}}{N_1!N_2!hL}\left(\frac{x}{L}\right)^{N_1}\left(1-\frac{x}{L}\right)^{N_2}e^{-\beta\frac{p^2}{2m_p}}.
\end{eqnarray}
These expressions describe two non-homogeneous gases and a fluctuating piston, as shown in Fig. \ref{fig:1} for two representative cases. Only for $N_1,N_2\to \infty$ (regardless the mass of the piston) are the two gases spatially homogeneous, with the position of the piston fixed (although with a fluctuating velocity). 

Despite the previous results, the system is spatially homogeneous at equilibrium, namely the probability of finding any particle (including the piston) in a given position is uniform. This can be seen from the density of particles, defined as
\begin{eqnarray}
  && n_i(\mathbf r,t)\equiv N_i \int d\mathbf p\ \rho_i(\mathbf r,\mathbf p, t), \qquad i=1,2, \\
  \label{eq:den1}
  && n_p(x,t)\equiv \int d p\ \rho_p(x,p, t),
\end{eqnarray}
for the particles on the left ($i=1$), the particles on the right ($i=2$), and the piston. At thermal equilibrium, each of the quantities depends on $x$, however it is easily seen that the global density $n$ is spatially uniform
\begin{equation}
n(x)\equiv  n_1(x)+n_2(x)+n_p(x)=\frac{N+1}{LS}.
\end{equation}

\begin{figure}[!h]
  \centering
  \includegraphics[width=.45\textwidth]{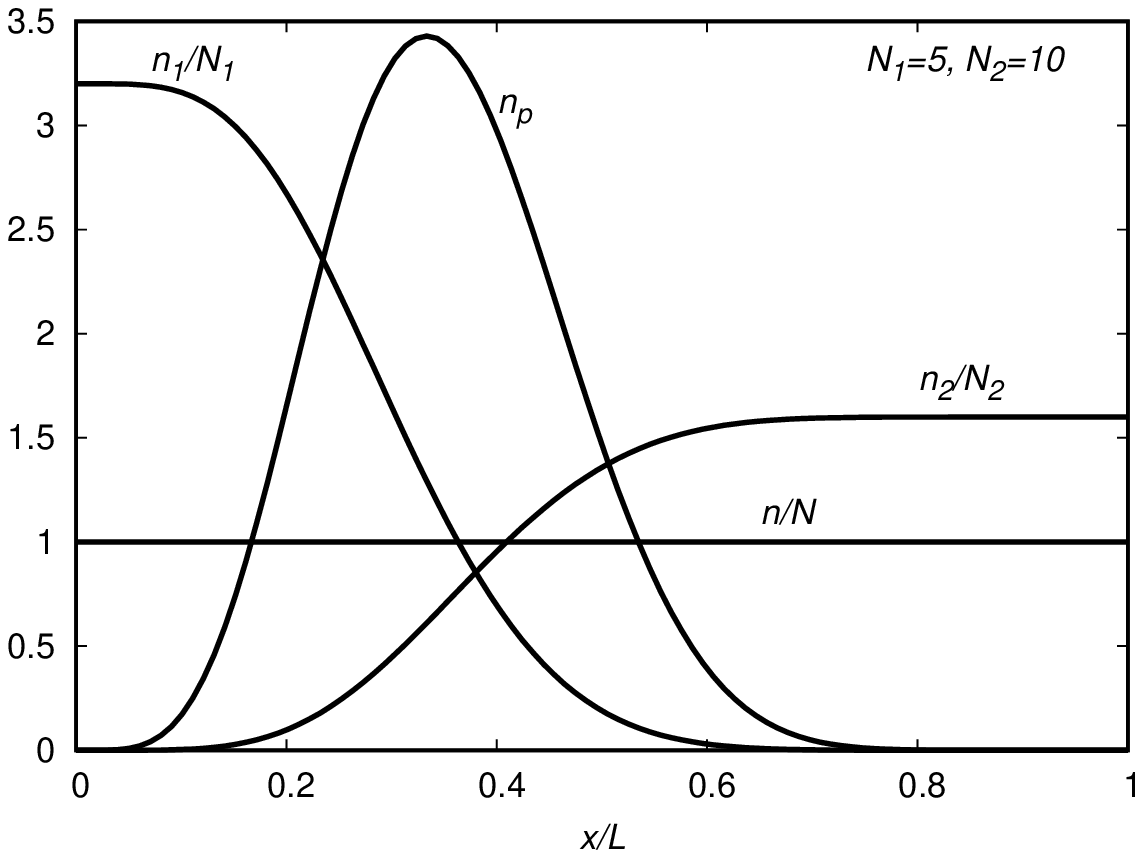}
  \includegraphics[width=.45\textwidth]{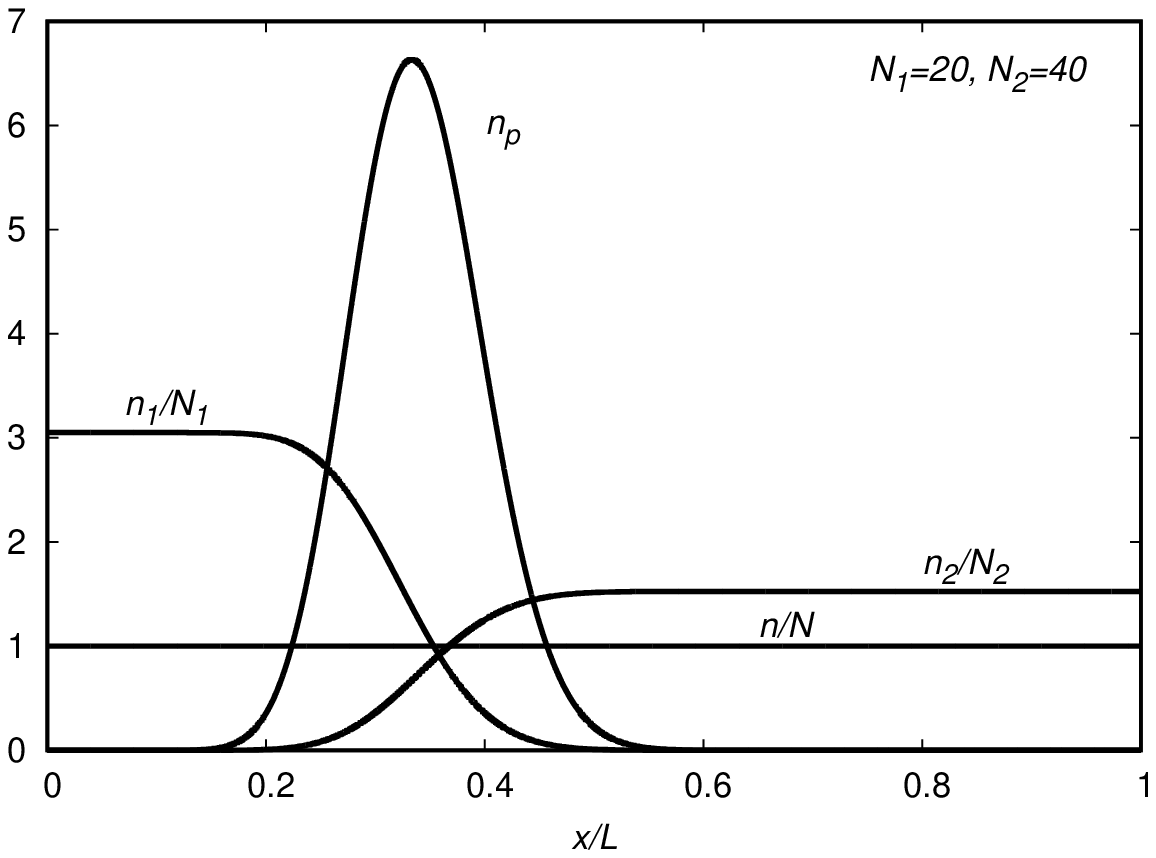}
  \caption{Probability densities for particles and the piston at equilibrium for different number of particles.}
  \label{fig:2}
\end{figure}

Similar arguments allow us to show that the pressure is  is $\frac{N+1}{LS\beta}$ along the $X$-axis and $\frac{N+1}{LS\beta}$ along any other normal direction. The calculation can be made using the partition function (then we should change the volume by fixing $S$ in the first case and fixing $L$ in the second one) or by computing the net flux of linear momentum across an imaginary $(d-1)$-surface. 

\subsection{Conditional probabilities}

As already shown, the fluctuations of the position of the piston make the probability densities for the gases to be non-homogenous at thermal equilibrium. Interestingly, the conditional probability densities of the gases to a position of the piston, $\rho_i(\mathbf r,\mathbf p|x_p)$ for $i=1,2$, are spatially homogenous at equilibrium. Using the definition of the conditional probabilities, and neglecting volume exclusion effects, we have
\begin{eqnarray}
 \nonumber 
  \rho_1(\mathbf r,\mathbf p|x_p)\equiv &&\frac{\rho_1(\mathbf r,x_p,\mathbf p)}{n_p(x_p)} \\ = \nonumber  &&\int_{x_i\le x_p}\prod_{i=1}^{N_1-1} d\mathbf r_id\mathbf p_i\int_{x_j\ge x_p}\prod_{j=N_1+1}^{N} d\mathbf r_jd\mathbf p_jdp_p\rho(\mathbf R,x_p;\mathbf P,p_p) \\ = && \frac{\lambda_1^d}{h^dSx_p}e^{-\beta \frac{p^2}{2m_1}}, \qquad x\le x_p,
\end{eqnarray}
and 
\begin{equation}
  \rho_2(\mathbf r,\mathbf p|x_p)=\frac{\lambda_2^d}{h^dS(L-x_p)}e^{-\beta \frac{p^2}{2m_2}}, \qquad x\ge x_p.
\end{equation}

There are two remarkable aspects of the previous results. On the one hand, the conditional distributions characterize the equilibrium state as if the piston where fixed. On the other hand, at thermal equilibrium the piston always ``feels'' homogeneous gases. The latter implies that we can also compute the equilibrium probability density of the piston as if it where in equilibrium under the force $F(x_p)$ exerted by the gases. Namely, at equilibrium, 
\begin{equation}
  \begin{split}
    F&=\int_{\frac{p_x}{m_1}>\frac{p_p}{m_p}}d\mathbf r_\perp d\mathbf p dp_p\ \left|\frac{p_x}{m_1}-\frac{p_p}{m_p}\right| \rho_1((x_p,\mathbf r_\perp),\mathbf p|x_p)\frac{\lambda_{p}}{h}e^{-\beta\frac{p_p^2}{2m_p}}\Delta^-p\\ &  -\int_{\frac{p_x}{m_2}<\frac{p_p}{m_p}}d\mathbf r_\perp d\mathbf p dp_p\ \left|\frac{p_p}{m_p}-\frac{p_x}{m_2}\right| \rho_2((x_p,\mathbf r_\perp),\mathbf p|x_p)\frac{\lambda_{p}}{h}e^{-\beta\frac{p^2}{2m_p}}\Delta^+p,
  \end{split}
\end{equation}
where 
\begin{equation}
  \label{eq:19}
  \begin{split}
    \Delta^-p=& \frac{2m_1m_p}{m_1+m_p}\left(\frac{p_x}{m_1}-\frac{p_p}{m_p}\right);  \\
    \Delta^+p=& \frac{2m_2m_p}{m_2+m_p}\left(\frac{p_p}{m_p}-\frac{p_x}{m_2}\right),
  \end{split}
\end{equation}
are the change of the linear momentum of the piston due to collisions with the gases. Operating, 
\begin{eqnarray}
  \nonumber
  F=&&\frac{2m_1m_p}{m_1+m_p}\frac{N_1\lambda_1^d\lambda_{p}}{h^{d+1}x_p}\int_{\frac{p_x}{m_1}>\frac{p_p}{m_p}} d\mathbf p dp_p\ \left(\frac{p_x}{m_1}-\frac{p_p}{m_p}\right)^2 e^{-\beta\left( \frac{p^2}{2m_1}+\frac{p_p^2}{2m_p}\right)} \\ \nonumber && -\frac{2m_2m_p}{m_2+m_p}\frac{N_2\lambda_2^d\lambda_{p}}{h^{d+1}(L-x_p)}\int_{\frac{p_x}{m_2}<\frac{p_p}{m_p}} d\mathbf p dp_p\ \left(\frac{p_p}{m_p}-\frac{p_x}{m_2}\right)^2 e^{-\beta\left( \frac{p^2}{2m_2}+\frac{p_p^2}{2m_p}\right)} \\ =&&\frac{1}{\beta}\left[\frac{N_1}{x_p}-\frac{N_2}{(L-x_p)}\right].
\end{eqnarray}
The associated ``effective'' potential $U(x)$ verifies $F(x)=-\frac{dU(x)}{dx}$ and is, up to an additive constant,
\begin{equation}
  U(x)=-\frac{1}{\beta}\ln\left[x^{N_1}(L-x)^{N_2}\right].
\end{equation}
Hence, the desired probability density of the piston is
\begin{equation}
  \rho_p(x,p)=\frac{1}{\mathcal Z_p}\exp\left[-\beta\left(\frac{p^2}{2m_p}+U(x)\right)\right]=\frac{x^{N_1}(L-x)^{N_2}}{\mathcal Z_p}e^{-\beta\frac{p^2}{2m_p}},
\end{equation}
with $\mathcal Z_p$ a normalization constant. The latter expression of $\rho_p$ coincides with Eq. \eqref{eq:rhop}.

\section{Kinetic description\label{sec:3}}

As shown in the previous section, even at equilibrium, there are spatial correlations between the gases and the piston. These correlations should be kept in order to have an accurate description of a finite, out-of-equilibrium system. Moreover, a easier characterization of the thermal equilibrium of the gases is possible using the conditional distributions. For these reasons, in this section we derive equations for the conditional distribution functions $f_i(\mathbf r,\mathbf v,t|x_p)$ for $i=1,2$ of the gases, as well as for the distribution function of the piston $f_p(x_p,v_p,t)$. 

\subsection{Preliminary definitions}

The distribution function of the gases $f_i(\mathbf r,\mathbf v,t)$ are defined such as $f_i(\mathbf r,\mathbf v,t)d\mathbf rd\mathbf v$ is the mean number of particles on the left ($i=1$) or on the right ($i=2$) with position between $\mathbf r$ and $\mathbf r+d\mathbf r$ and velocity between $\mathbf v$ and $\mathbf v+d\mathbf v$ at time $t$. They are related with the probability densities $\rho_i(\mathbf r,\mathbf p,t)$ as 
\begin{equation}
  f_i(\mathbf r,\mathbf v,t)=N_im_i^{d}\rho_i(\mathbf r,\mathbf v/m_i,t), \quad i=1,2.
\end{equation}
The distribution function of the piston $f_p(x_p,v_p,t)$ is the probability density of finding the piston with position $x_p$ and velocity $v_p$ at time $t$, now
\begin{equation}
  f_p(x_p, v_p,t)=m_p\rho_p(x_p,v_p/m_p,t).
\end{equation}

We define the conditional distribution of the gases as
\begin{equation}
  \label{eq:cond}
  f_i(\mathbf r,\mathbf v,t|x_p)\equiv \frac{f_i(\mathbf r,\mathbf v,x_p,t)}{n_p(x_p,t)}, \quad i=1,2,
\end{equation}
where $f_i(\mathbf r,\mathbf v,x_p,t)$ is the joint probability distribution for a particle of gas $i$ with position $\mathbf r$ and velocity $\mathbf v$ and the piston with position $x_p$, regardless its velocity, at time $t$. It is convenient to express the new probability density in terms of the more general joint probability $f_{ip}(\mathbf r,\mathbf v,x_p,v_p,t)$ where the velocity of the piston is also specified,  
\begin{equation}
  f_{i}(\mathbf r,\mathbf v,x_p,t)\equiv \int dv_p\ f_{ip}(\mathbf r,\mathbf v,x_p,v_p,t).
\end{equation}
The other quantity in Eq. \eqref{eq:cond} is the density of the piston $n_p(x_p,t)$ already defined in Eq. \eqref{eq:den1}, now given in terms of $f_p$ as 
\begin{equation}
  n_p(x_p,t)=\int dv_p \ f_p(x_p,v_p,t). 
\end{equation}

Finally, it is convenient for forthcoming discussions to define the conditional probability distribution of a position and a velocity of the gases to a position and velocity of the piston as
\begin{equation}
  f_i(\mathbf r,\mathbf v,t|x_p,v_p)\equiv \frac{f_{ip}(\mathbf r,\mathbf v,x_p,v_p,t)}{f_p(x_p,v_p,t)}, \quad i=1,2.
\end{equation}
It is forth observing that $f_i(\mathbf r,\mathbf v,t|x_p)\ne \int f_i(\mathbf r,\mathbf v,t|x_p,v_p)$ in general, due to velocity correlations. An exception is when the system is at thermal equilibrium.

\subsection{Set of Boltzmann-like kinetic equations}

\subsubsection{Kinetic equation for the piston}
If we were to proceed as usual, we should write down a set of equations for the distribution functions, of the gases and the piston, from the Liouville equation. This would result in a non-closed set of equations, since higher order distribution function would appear. The difference in the present case with respect to other problems is that the conventional molecular chaos hypothesis does not work, and new approximations are needed. 

Take first the kinetic equation for $f_p(x_p,v_p,t)$, resulting form integrating the Liouville equation over all variables except the position and velocity of the system,
\begin{equation}
  \label{eq:pist}
  \partial_t f_p+v_p \partial_{x_p}f_p=J_{p1}[v_p|f_1,f_p]+J_{p2}[v_p|f_2,f_p].
\end{equation}
It is nothing but a probabilistic balance equation, where the lhs accounts for the free-of-collision motion of the piston and the rhs accounts for the collisions with the gases. The latter includes the collision operators, 
\begin{eqnarray}
  \nonumber J_{p1}=&&\int d\mathbf r_\perp\int_{v_x^*>v_p^*}d\mathbf v\ (v_x^*-v_p^*)f_{1p}((x_p,\mathbf r_\perp),\mathbf v^*,x_p,v_p^*,t)\\ && -\int d\mathbf r_\perp\int_{v_x>v_p}d\mathbf v\ (v_x-v_p)f_{1p}((x_p,\mathbf r_\perp),\mathbf v,x_p,v_p,t), \\ \nonumber J_{p2}=&&\int d\mathbf r_\perp\int_{v_x^*<v_p^*}d\mathbf v\ (v_p^*-v_x^*)f_{2p}((x_p,\mathbf r_\perp),\mathbf v^*,x_p,v_p^*,t)\\ && -\int d\mathbf r_\perp\int_{v_x<v_p}d\mathbf v\ (v_p-v_x)f_{2p}((x_p,\mathbf r_\perp),\mathbf v,x_p,v_p,t),
\end{eqnarray}
which are functions of $v_p$ and functionals of the distribution functions. The asterisk on the velocities indicates precollisional values,  
\begin{equation}
  \begin{split}
    v_x^*\equiv & v_x-\frac{2m_p}{m_i+m_p}(v_x-v_p), \\
    \mathbf v_\perp^*\equiv & \mathbf v_\perp, \\
    v_p^*\equiv & v_p-\frac{2m_i}{m_i+m_p}(v_p-v_x),
  \end{split}
\end{equation}
for $i=1$ or $2$. The functional dependence of the operator can be made more clear if we replace the joint probabilities using the following relation
\begin{equation}
  f_{ip}((x_p,\mathbf r_\perp),\mathbf v,x_p,v_p,t)=f_{i}((x_p,\mathbf r_\perp),\mathbf v,t|x_p,v_p) f_p(x_p,v_p,t).
\end{equation}

\paragraph{Assumption 1.} The usual molecular chaos hypothesis for precollisions is translated in this work to
\begin{equation}
  f_{i}((x_p,\mathbf r_\perp),\mathbf v,t|x_p,v_p)\simeq f_{i}((x_p,\mathbf r_\perp),\mathbf v,t|x_p),
\end{equation}
for precollisions. That is, we keep the correlations in the position but remove precollisional velocity correlations.

Observe that the functional dependence of the collision operators are evaluated when a particle is at the position of the piston, dissregarding the diameter of the particle. Since we are interested in a mesoscopic description where the relevant spatial variation of the distribution function occur over a mean free path of the particles, the simplification is expected to be accurate for the gases dilute enough.

\subsubsection{Kinetic equations for the gases}

To complete the kinetic description, we need equations for the conditional probabilities $f_i(\mathbf r,\mathbf v,t|x_p)$. For that purpose, and according to the definition of Eq. \eqref{eq:cond}, we need equations for $n_p(x_p,t)$ and $f_{ip}(\mathbf r,\mathbf v,x_p,v_p,t)$.

Integrating Eq. \eqref{eq:pist}, we get 
\begin{eqnarray}
  \label{eq:np}
  \nonumber 
    \partial_tn_p(x_p,t)= && \int dv_p \partial_tf_p(x_p,v_p,t)=\int dv_p\left(-v_p\partial_{x_p}f_p+J_{p1}+J_{p2}\right) \\ =&&
    -\partial_{x_p}\left[u_p(x_p,t)n_p(x_p,t)\right],
\end{eqnarray}
where 
\begin{equation}
  u_p(x_p,t)=\frac{1}{n_p(x_p,t)}\int dv_p\ v_p f_p(x_p,v_p,t)
\end{equation}
is the mean velocity of the piston. In Eq.~\eqref{eq:np} we have used the conservation of the number of particles in collisions,
\begin{equation}
    \int dv_p J_{p1}=\int dv_p J_{p2}=0.
\end{equation}

The kinetic equation for the joint probability $f_{1p}(\mathbf r,\mathbf v,x_p,v_p,t)$, resulting from the Liouville equation after a proper integration, is
\begin{equation}
  \label{eq:f1p}
  \partial_t f_{1p}=-(\mathbf v\cdot \partial_{\mathbf r}+v_p\partial_{x_p})f_{1p}+J_{1p,11'}+J_{1p,p1'}+J_{1p,p2},
\end{equation}
where the collision operators take into account the different mechanisms the distribution function $f_{1p}$ can change through collisions. For the operator $J_{1p,11'}$ the collision is between two particles on the left of the piston, for $J_{1p,p1'}$ the collision is between the piston and a second particle on the left, and $J_{1p,p2}$ accounts for collisions between the piston and a particle on the right. These collision terms involve three-body distribution functions. There is still one kind of collision to consier, involving a particle with position and velocity given by the argument of $f_{1p}$ and the piston, when they are in contact ($x=x_p)$ and about to collide ($v_x\ge v_p$). In this case, the usual procedure is to impose a boundary condition to $f_{1p}$, which is a consequence of the conservation of probability:
\begin{equation}
  \label{eq:f1pbc}
  (v_p'-v_x')f_{1p}((x_p,\mathbf r_\perp),\mathbf v',x_p,v'_p,t)=(v_x-v_p)f_{1p}((x_p,\mathbf r_\perp),\mathbf v,x_p,v_p,t),
\end{equation}
for  $v_x\ge v_p$. The primed variables are given by the collision rule in Eqs. \eqref{eq:coll}-\eqref{eq:coll2}. For the gas on the right, an analogous equation and boundary condition result.

Before proceeding with the derivation of the equation for $f_i(\mathbf r,\mathbf v,t|x_p)$, it is worth focusing on some consequences of the boundary condition in Eq. \eqref{eq:f1pbc}. If we integrate it over velocities for $v_x\ge v_p$, after a change of variable and taking into account that $dv_xdv_p=dv_x'dv_p'$, we get
\begin{equation}
  \label{eq:bvel}
  n_p(x_p,t)n_i(x_p,t)u_p(x_p,t)=n_p(x_p,t)n_i(x_p,t)u_{i,x}(x_p,t), \quad i=1,2,
\end{equation}
with the mean velocity of the gases $\mathbf u_i(\mathbf r,t)$ ($i=1,2$) defined as
\begin{equation}
  n_i(\mathbf r,t)\mathbf u_i(\mathbf r,t)\equiv \int d\mathbf v\ \mathbf v f_i(\mathbf r,\mathbf v,t), \quad i=1,2.
\end{equation}
For nonzero densities, the velocity of the gases and the piston in contact coincide.

Integrating Eq. \eqref{eq:f1p} over the velocity of the piston, we have
\begin{equation}
  \partial_t \int dv_p f_{1p}= -\mathbf v\cdot \partial_{\mathbf r}\int dv_p f_{1p}-\partial_{x_p}\int dv_p\ v_p f_{1p}+\int dv_p\ J_{1p,11'},
\end{equation}
where we have used the fact that the collision operators conserve the number of particles. Using last equation and Eq. \eqref{eq:np} with the temporal derivative of Eq. \eqref{eq:cond}, after some manipulations, we arrive at 
\begin{eqnarray}
  \nonumber
  \partial_t f_1(\mathbf r,\mathbf v,t|x_p)=&&f_1(\mathbf r,\mathbf v,t|x_p)\frac{\partial_{x_p}\left[u_p(x_p,t)n_p(x_p,t)\right]}{n_p(x_p,t)}-\frac{1}{n_p(x_p,t)}\mathbf v\cdot \partial_{\mathbf r}\int dv_p f_{1p}\\ && -\frac{1}{n_p(x_p,t)}\partial_{x_p}\int dv_p\ v_p f_{1p}+\frac{1}{n_p(x_p,t)}\int dv_p\ J_{1p,11'},
\end{eqnarray}
which is not a closed equation yet.

\paragraph{Assumptions 2 and 3.} In order to close the equation we include additional approximations  that go beyond the standard molecular chaos hypothesis, namely
\begin{eqnarray}
  \label{eq:ap1}
  \nonumber 
  \int dv_p\ v_pf_{1p}(\mathbf r,\mathbf v,x_p,v_p,t)=&& \int dv_p\ v_pf_{1p}(\mathbf r,\mathbf v,t|x_p,v_p)f_p(x_p,v_p,t)\\ \simeq&& \nonumber f_{1p}(\mathbf r,\mathbf v,t|x_p)\int dv_p\ v_pf_p(x_p,v_p,t)\\= && n_p(x_p,t)u_p(x_p,t)f_{1p}(\mathbf r,\mathbf v,t|x_p), 
\end{eqnarray}
and
\begin{equation}
  \label{eq:ap2}
  \int dv_p\ J_{1p,11'}\simeq  J_{11} \int dv_p\ f_{1p}(\mathbf r,\mathbf v,t|x_p)=n_p(x_p,t)J_{11},
\end{equation}
with $J_{11}$ the collision operator of two particles on the left, which is a function of $\mathbf v$ and a functional of $f_1(\mathbf r,\mathbf v,t|x_p)$.  For hard-sphere collisions, the latter approximation is
\begin{eqnarray}
  \nonumber 
\int dv_p\ J_{1p,11'}    &&=\int dv_p\ f_p(x_p,v_p,t) \int d\mathbf v_2 d\hat{\boldsymbol{\sigma}}\ \Theta [(\mathbf{v}-\mathbf{v}_2)\cdot \hat{\boldsymbol{\sigma}}](\mathbf{v}-\mathbf{v}_2)\cdot \hat{\boldsymbol{\sigma}} \\ && \nonumber \quad \times  (b^{-1}-1) f_{1}(\mathbf r,\mathbf v,t|x_p,v_p)f_{1}(\mathbf r,\mathbf v_2,t|x_p,v_p) \\ \nonumber &&\simeq \int dv_p\ f_p(x_p,v_p,t) \int d\mathbf v_2 d\hat{\boldsymbol{\sigma}}\ \Theta [(\mathbf{v}-\mathbf{v}_2)\cdot \hat{\boldsymbol{\sigma}}](\mathbf{v}-\mathbf{v}_2)\cdot \hat{\boldsymbol{\sigma}} \\ && \nonumber \quad \times  (b^{-1}-1) f_{1}(\mathbf r,\mathbf v,t|x_p)f_{1}(\mathbf r,\mathbf v_2,t|x_p)\\ &&= n_p(x_p,t)J_{11},
\end{eqnarray}
where $b^{-1}$ is the restitution operator, which replaces the velocities of the colliding particles by their precollisional values, and
\begin{eqnarray}
  \nonumber
  J_{11}= \int d\mathbf v_2 d\hat{\boldsymbol{\sigma}}\ && \Theta [(\mathbf{v}-\mathbf{v}_2)\cdot \hat{\boldsymbol{\sigma}}](\mathbf{v}-\mathbf{v}_2)\cdot \hat{\boldsymbol{\sigma}} \\ && \times (b^{-1}-1) f_{1}(\mathbf r,\mathbf v,t|x_p)f_{1}(\mathbf r,\mathbf v_2,t|x_p).
\end{eqnarray}

Finally, we arrive at a closed equation for $f_1(\mathbf r,\mathbf v,t|x_p)$,
\begin{equation}
  \label{eq:eqf1}
  \partial_t f_1(\mathbf r,\mathbf v,t|x_p) \simeq-\mathbf v\cdot \partial_{\mathbf r}f_1(\mathbf r,\mathbf v,t|x_p)-u_p(x_p,t)\partial_{x_p}f_1(\mathbf r,\mathbf v,t|x_p) +J_{11}.
\end{equation}
With similar approximation, we also obtain the kinetic equation for the gas on the right of the piston,
\begin{equation}
  \label{eq:eqf2}
  \partial_t f_2(\mathbf r,\mathbf v,t|x_p) \simeq-\mathbf v\cdot \partial_{\mathbf r}f_2(\mathbf r,\mathbf v,t|x_p)-u_p(x_p,t)\partial_{x_p}f_2(\mathbf r,\mathbf v,t|x_p) +J_{22},
\end{equation}
with
\begin{eqnarray}
  \nonumber
  J_{22}= \int d\mathbf v_2 d\hat{\boldsymbol{\sigma}}\ && \Theta [(\mathbf{v}-\mathbf{v}_2)\cdot \hat{\boldsymbol{\sigma}}](\mathbf{v}-\mathbf{v}_2)\cdot \hat{\boldsymbol{\sigma}} \\ && \times (b^{-1}-1) f_{2}(\mathbf r,\mathbf v,t|x_p)f_{2}(\mathbf r,\mathbf v_2,t|x_p).
\end{eqnarray}

For the boundary conditions, we have to differentiate from particles in contact with the walls of the cylinder, denoted by $\partial V$, and the piston. In the first case, the conservation of probability imposes 
\begin{equation}
  \label{eq:bcdv}
    f_i(\mathbf r,\mathbf v,t|x_p)=f_i(\mathbf r,\mathbf v',t|x_p), \quad \mathbf r\in \partial V, \quad i=1,2,
  \end{equation}
  for $\mathbf v\cdot \hat{\boldsymbol{n}}=-\mathbf v'\cdot \hat{\boldsymbol{n}}$, where $\hat{\boldsymbol{n}}$ is a unit vector normal to $\partial V$.

  \paragraph{Assumption 4.} For the boundary conditions at the position of the piston, different approximations are possible. We take the one resulting from Eq. \eqref{eq:f1pbc}, after integration over the velocity of the piston and assuming $f_1(\mathbf r,\mathbf v,t|x_p,v_p) \simeq f_1(\mathbf r,\mathbf v,t|x_p) $ inside the integrals,
\begin{eqnarray}
  \label{eq:bc1}
\nonumber 
  && \int_{v_x>v_p} dv_p\ (v_p'-v_x')f_{1}((x_p,\mathbf r_\perp),\mathbf v',t|x_p)f_p(x_p,v_p',t) \\ && \qquad \simeq  \int_{v_x>v_p}\ (v_x-v_p)f_{1}((x_p,\mathbf r_\perp),\mathbf v,t|x_p)f_p(x_p,v_p,t).
\end{eqnarray}
and
\begin{eqnarray}
  \label{eq:bc2}
\nonumber 
  && \int_{v_x<v_p} dv_p\ (v_p'-v_x')f_{2}((x_p,\mathbf r_\perp),\mathbf v',t|x_p)f_p(x_p,v_p',t) \\ && \qquad \simeq  \int_{v_x<v_p}\ (v_x-v_p)f_{2}((x_p,\mathbf r_\perp),\mathbf v,t|x_p)f_p(x_p,v_p,t).
\end{eqnarray}
These approximate boundary conditions have the advantage, if compared to others, to conserve the exact relation of Eq. \eqref{eq:bvel}, which is of crucial importance for the conservation of particles, as seen later.  

\subsection{Canonical equilibrium}

The proposed system of equations \eqref{eq:pist},\eqref{eq:eqf1}-\eqref{eq:eqf2} with boundary conditions \eqref{eq:bc1}-\eqref{eq:bc2} has the canonical equilibrium as a solution, given by
\begin{eqnarray}
  \label{eq:f1eq}
  f_1((x,\mathbf r_\perp),\mathbf v,t|x_p)=&&\frac{N_1}{x_pS}\frac{m_1^d\lambda_1^d}{h^d}e^{-\beta\frac{1}{2}m_1v^2}, \quad x\le x_p, \\
  \label{eq:f2eq}
    f_2((x,\mathbf r_\perp),\mathbf v,t|x_p)=&&\frac{N_1}{(L-x_p)S}\frac{m_2^d\lambda_2^d}{h^d}e^{-\beta\frac{1}{2}m_2v^2}, \quad x\ge x_p, \\
  \label{eq:fpeq}
  f_p(x_p,v_p,t)=&&n_p(x_p)\frac{m_p\lambda_p}{h}e^{-\beta\frac{1}{2}m_pv_p^2},
\end{eqnarray}
with
\begin{equation}
  \label{eq:nxp}
n_p(x_p)=\frac{(N+1)!}{N_1!N_2!L}\left(\frac{x}{L}\right)^{N_1}\left(1-\frac{x}{L}\right)^{N_2}.
\end{equation}

To demonstrate the latter assertion, take $J_{p1}$ at canonical equilibrium. Since $v_x^*-v_p^*=-(v_x-v_p)$ and $m_1v_x^{*2}+m_pv_p^{*2}=m_1v_x^{2}+m_pv_p^{2}$, after integrating over $\mathbf v_\perp$, it is
\begin{eqnarray}
  \nonumber
  J_{p1}[v_p|f_p,f_1]=&&-\frac{N_1m_1\lambda_1m_p\lambda_p}{x_ph^2}\rho_p(x_p)\int_{v_x<v_p}dv_x\ (v_x-v_p)e^{-\beta\frac{1}{2}m_pv_p^2-\beta\frac{1}{2}m_1v_x^2}\\   \nonumber
 &&  -\frac{N_1m_1\lambda_1m_p\lambda_p}{x_ph^2}\rho_p(x_p)\int_{v_x>v_p}dv_x\ (v_x-v_p)e^{-\beta\frac{1}{2}m_pv_p^2-\beta\frac{1}{2}m_1v_x^2} \\ =&&\frac{N_1m_p\lambda_p}{x_ph}n_p(x_p)v_pe^{-\beta\frac{1}{2}m_pv_p^2}=\frac{N_1}{x_p}v_pf_p(x_p,v_p,t).
\end{eqnarray}
Similarly, 
\begin{equation}
  \label{eq:31}
  J_{p2}[v_p|f_p,f_2]=-\frac{N_2}{L-x_p}v_pf_p(x_p,v_p,t).
\end{equation}
Hence,
\begin{equation}
  \frac{d}{dx_p}n_p(x_p)=\frac{N_1}{x_p}-\frac{N_2}{L-x_p}v_pf_p(x_p,v_p,t),
\end{equation}
which has expression \eqref{eq:nxp} as the unique normalized solution. 

The considerations for the conditional probabilities are straightforward. Since $J_{11}=J_{22}=0$ for gaussian distributions, Eqs. \eqref{eq:eqf1} and \eqref{eq:eqf2} clearly support the equilibrium solutions. 

\subsection{Some comments and remarks}

The equations for the distribution function of the piston \eqref{eq:pist} and the conditional distribution functions for the gases \eqref{eq:eqf1}-\eqref{eq:eqf2} constitute one of the main results of the present work. The chose of the conditional distribution as the function to describe the gases has been motivated by the analysis of the canonical equilibrium, since there we see the need of including spatial correlations between gases and the piston, and the equation of the piston where the conditional probabilities emerge in a natural way. 
 
The derivation of the kinetic equations used several approximations, namely the molecular chaos for inter-particle and piston-particle collisions (assumption 1), and a generalization of molecular chaos for particle-piston collisions (assumptions 2-4). The molecular chaos hypothesis neglects pre-collisional velocity correlations, as usual when dealing with the Boltzmann equations. The generalized molecular chaos hypothesis goes one step further by also removing ``some'' post-collisional velocity correlations, as specified by Eqs. \eqref{eq:ap1}-\eqref{eq:ap2} and \eqref{eq:bc1}-\eqref{eq:bc2}.  The approximate expression of Eqs. \eqref{eq:ap1}-\eqref{eq:ap2} are expected to be correct as far as $f_i(\mathbf r,\mathbf v,t|v_p,x_p)\simeq f_i(\mathbf r,\mathbf v,t|x_p)$ inside the integrals. This, in turn, is expected to be true in a general situation with $x$ not very close to $x_p$, since collisions between particles and the piston, as well as the collisions among particles, tend to destroy the velocity correlations. When $x\simeq x_p$, which is the case of Eqs. \eqref{eq:bc1}-\eqref{eq:bc2}, the removal of correlations is not that obvious. In fact, the boundary conditions  \eqref{eq:bc1}-\eqref{eq:bc2} are strictly wrong. However, if the section of the piston $S$ is big enough $S/\sigma_i^{d-1}\gg 1$, we could find a regime of dilute gases where the postcollisional velocity correlations decay to zero on a distance much smaller than the mean free path of the particles, meaning that the boundary conditions become correct in the mesoscale. An exact analysis of this considerations needs a formulation with the functions $f_i(\mathbf r,\mathbf v,t|v_p,x_p)$, which is beyond the scope of the present work. 

Observe that the structure of the kinetic equations for the conditional distributions of the gases  is different from the one expected from naive considerations, as terms proportional to the mean velocity of the piston appear. Although the presence of these terms can be justified by means of physical considerations, i.e they are important in order to compute the work done by the piston correctly, they can compromise some key properties of the equation. Namely, for a given physical initial condition, the kinetic equations for the conditional distribution functions should provide positive and normalized solutions for all later times. Both aspects could be indirectly demonstrated from the equation of $f_{ip}$ and $n_p$ or directly. As an example, we show in the sequel that the kinetic equations \eqref{eq:pist}-\eqref{eq:eqf1}-\eqref{eq:eqf2} preserve the normalization of distribution functions.  

Consider first the equation of the density of the piston Eq. \eqref{eq:np}. Integrating over $x_p\in(0,L)$, using the divergence theorem, and the fact that the piston can not leave the system, $n_p(0,t)u_p(0,t)=n_p(L,t)u_p(L,t)=0$, it turns out that $\int dx_p\ n_p(x_p,t)$ is a conserved quantity. Hence, the normalization of $f_p$ is ensured, provided it is initially normalized. 

Consider now the gas on the left. After integration of Eq. \eqref{eq:eqf1} over $\mathbf v$, we face several terms:
\begin{equation}
  \int d\mathbf v\ \partial_t f_1(\mathbf r,\mathbf v,t|x_p)=\partial_tn_1(\mathbf r,t|x_p),
\end{equation}
with
\begin{equation}
  n_1(\mathbf r,t|x_p)\equiv \int d\mathbf v\ f_1(\mathbf r,\mathbf v,t|x_p)
\end{equation}
being the conditional density;
\begin{equation}
  \int d\mathbf v\ \mathbf v\cdot \partial_{\mathbf r}f_1(\mathbf r,\mathbf v,t|x_p)=\partial_{\mathbf r}\cdot[n_1(\mathbf r,t|x_p)\mathbf u_1(\mathbf r,t|x_p)],
\end{equation}
with the conditional velocity defined as 
\begin{equation}
  \mathbf u_1(\mathbf r,t|x_p)\equiv \frac{1}{n_1(\mathbf r,t|x_p)}\int d\mathbf v \ \mathbf v f_1(\mathbf r,\mathbf v,t|x_p);
\end{equation}
\begin{equation}
  \int d\mathbf v\ u_p(x_p,t)\partial_{x_p}f_1(\mathbf r,\mathbf v,t|x_p)=u_p(x_p,t)\partial_{x_p}n_1(\mathbf r,t|x_p);
\end{equation}
and
\begin{equation}
 \int d\mathbf v\ J_{11}=0.
\end{equation}
Hence, the balance equation for the conditional density is
\begin{equation}
  \partial_{t}n_1(\mathbf r,t|x_p)=-\partial_{\mathbf r}\cdot[n_1(\mathbf r,t|x_p)\mathbf u_1(\mathbf r,t|x_p)]-u_p(x_p,t)\partial_{x_p}n_1(\mathbf r,t|x_p),
\end{equation}
and a similar one for the conditional density of the gas on the right of the piston $n_2(\mathbf r,t|x_p)$. Integrating over $x\in(0,x_p)$ and all possible values of $\mathbf r_\perp$:
\begin{eqnarray}
  \nonumber
  \partial_t N_1(t|x_p)=&& -\left. \int d\mathbf r_\perp\ n_1(\mathbf r,t|x_p) u_{1,x}(\mathbf r,t|x_p)\right|_{x=x_p} \\ &&-u_p(x_p,t)\int d\mathbf r_\perp \int_0^{x_p}dx \partial_{x_p}n_1(\mathbf r,t|x_p),
\end{eqnarray}
with 
\begin{equation}
  N_1(t|x_p)\equiv \int d\mathbf r\ n_1(\mathbf r,t|x_p),
\end{equation}
We have used the condition 
\begin{equation}
  \mathbf u_1(x\in \partial V,t|x_p)\cdot \hat{\boldsymbol{n}}=0,
\end{equation}
with $\hat{\boldsymbol{n}}$ a unit vector normal to boundary of the cylinder $\partial V$, which is a direct consequence of the kinetic boundary condition \eqref{eq:bcdv}. The quantity $u_{1,x}$ is the $X$ component of $\mathbf u_1(\mathbf r,t|x_p)$. Now, 
\begin{equation}
  \int_0^{x_p}dx \partial_{x_p}n_1(\mathbf r,t|x_p)=\partial_{x_p}N_1(t|x_p)-n_1(x_p,t|x_p).
\end{equation}
Hence, 
\begin{eqnarray}
  \nonumber
  \partial_t N_1(x_p,t)=&&-\left. \int d\mathbf r_\perp\ n_1(\mathbf r,t|x_p) [u_{1,x}(\mathbf r,t|x_p) -u_p(x_p,t)]\right|_{x=x_p}\\ && -u_p(x_p,t)\partial_{x_p}N_1(t|x_p).
\end{eqnarray}
The first term on the rhs is zero, using the kinetic boundary condition at the piston \eqref{eq:bc1},
\begin{eqnarray}
  \nonumber
  && \int_{v_x>v_p} dv_xdv_p(v_x-v_p)f_1(x_p,v_x,t|x_p)f_p(x_p,v_p,t)\\ && +\int_{v_x>v_p} dv_xdv_p (v_x'-v_p')f_1(x_p,v_x',t|x_p)f_p(x_p,v_p',t)\simeq 0
\end{eqnarray}
which, making a change of variable on the second integral, turns
\begin{eqnarray}
  \nonumber
  && \int dv_xdv_p(v_x-v_p)f_1(x_p,v_x,t|x_p)f_p(x_p,v_p,t)\simeq 0 \\ \nonumber &&\Rightarrow n_1(x_p,t|x_p)n_p(x_p,t)(u_{1,x}-u_{p})\simeq 0\\ && \Rightarrow n_1(x_p,t|x_p)(u_{1,x}-u_{p})\simeq 0,
\end{eqnarray}
if $n_p(x_p,t)\ne 0$. Hence, the balance equation for $N_1$ results
\begin{equation}
  \partial_t N_1(t|x_p)=-u_p(x_p,t)\partial_{x_p}N_1(t|x_p).
\end{equation}
From the nature of this equation we infer that if the conditional distribution function for the gas on the left is normalized to $N_1$  for all allowed values of $x_p$, that is $N_1(0|x_p)=N_1$, then $N_1(t|x_p)=N_1$ for $t\ge 0$.

 Similar considerations allow us to conclude that the conditional distribution of the gas on the right of the piston is also normalized to $N_2$ for all times.

\section{$\mathcal{H}$-theorem\label{sec:4}}
In this section we demonstrate the $\mathcal H$-theorem for the system of kinetic equations derived along the previous section. It states that the $\mathcal H$ function, to be defined below, mush approach a limit where the distribution functions are of thermal equilibrium given by Eqs. \eqref{eq:f1eq}-\eqref{eq:fpeq}. We first prove $\mathcal H$ is a decreasing function, then that its time derivative is zero at thermal equilibrium, and finally that it is bounded from below by its value at equilibrium. 

\subsection{Definition of $\mathcal H$}
Following a recent work \cite{magabr18}, take the $\mathcal{H}$ function as 
\begin{equation}
  \mathcal H\equiv \int d\Gamma \rho_N(\Gamma,t)\ln\left[\frac{\rho_N(\Gamma,t)}{\rho_{01}^{N_1}\rho_{02}^{N_2}\rho_{0p}}\right],
\end{equation}
with $\rho_N$ the probability density of the whole system, $\Gamma\equiv (\mathbf R,x_p;\mathbf P,p_p)$, and $\rho_{0i}$ constants that make the expression dimensionless. Removing velocity correlations, as usual, it is
\begin{eqnarray}
  \nonumber 
  \rho_N(\mathbf R,x_p;\mathbf P,p_p,t)&&=\rho_N(\mathbf R;\mathbf P,t|p_p,x_p)\rho_p(x_p,p_p,t)\\ \nonumber && \simeq \rho_N(\mathbf R;\mathbf P,t|x_p)\rho_p(x_p,p_p,t)\\ && \simeq \prod_{i=1}^{N_1}\rho_1(\mathbf r_i,\mathbf p_i,t|x_p)\prod_{j=1}^{N_2}\rho_2(\mathbf r_j,\mathbf p_j,t|x_p)\rho_p(x_p,p_p,t),
\end{eqnarray}
and the $\mathcal H$-function turns
\begin{eqnarray}
  \nonumber 
  \mathcal H \simeq && N_1 \int d\mathbf rdx_pd\mathbf pdp_p\ \rho_1(\mathbf r,\mathbf p,t|x_p)\rho_p (x_p,p_p,t)\ln\left[\frac{\rho_1(\mathbf r,\mathbf p,t|x_p)}{\rho_{01}}\right]  \\  \nonumber && + N_2 \int d\mathbf rdx_pd\mathbf pdp_p\ \rho_2(\mathbf r,\mathbf p,t|x_p)\rho_p (x_p,p_p,t)\ln\left[\frac{\rho_2(\mathbf r,\mathbf p,t|x_p)}{\rho_{02}}\right] \\ && + \int dx_p dp_p\ \rho_p(x_p,p_p,t)\ln\left[\frac{\rho_p(x_p,p_p,t)}{\rho_{0p}}\right].
\end{eqnarray}
In terms of the distribution functions, 
\begin{eqnarray}
  \label{eq:etaf}
  \nonumber
  \mathcal H \simeq && \int d\mathbf rdx_pd\mathbf vdv_p\ f_1(\mathbf r,\mathbf v,t|x_p)f_p (x_p,v_p,t)\ln\left[\frac{f_1(\mathbf r,\mathbf v,t|x_p)}{f_{01}}\right]  \\ \nonumber && + \int d\mathbf rdx_pd\mathbf vdv_p\ f_2(\mathbf r,\mathbf v,t|x_p)f_p (x_p,v_p,t)\ln\left[\frac{f_2(\mathbf r,\mathbf v,t|x_p)}{f_{02}}\right] \\ && + \int dx_p dv_p\ f_p(x_p,v_p,t)\ln\left[\frac{f_p(x_p,v_p,t)}{f_{0p}}\right],
\end{eqnarray}
with the new constants $f_{0i}$ having the dimensions of their respective distribution functions.

\subsection{Proof of $\frac{d}{dt}\mathcal H \le 0$}
Now we compute the time derivative of $\mathcal H$ as given by its last expression. For the sake of simplicity on the notation, we omit the obvious dependence of the different functions.

The time derivative of the first term  of Eq. \eqref{eq:etaf} is 
\begin{eqnarray}
  \nonumber
    \frac{d}{dt}\int d\mathbf rdx_pd\mathbf v\ f_1n_p\ln\left(\frac{f_1}{f_{01}}\right)= && \int d\mathbf rdx_pd\mathbf v\ \left[1+\ln\left(\frac{f_1}{f_{01}}\right)\right]n_p\partial_tf_1 \\ && + \int d\mathbf rdx_pd\mathbf v\ f_1\ln\left(\frac{f_1}{f_{01}}\right)\partial_tn_p,
\end{eqnarray}
which is, after using the equations for $f_1$ and $n_p$ \eqref{eq:eqf1} and \eqref{eq:np}, 
\begin{eqnarray}
  \nonumber
  && \int d\mathbf rdx_pd\mathbf v\ \left[1+\ln\left(\frac{f_1}{f_{01}}\right)\right]n_p\left(-\mathbf v\cdot \partial_{\mathbf r}f_1-u_p\partial_{x_p}f_1+J_{11}\right) \\ &&- \int d\mathbf rdx_pd\mathbf v\ f_1\ln\left(\frac{f_1}{f_{01}}\right)\partial_{x_p}\left(u_pn_p\right).
\end{eqnarray}
After some algebra, 
\begin{eqnarray}
  && \left[1+\ln\left(\frac{f_1}{f_{01}}\right)\right]n_p \left(-\mathbf v\cdot \partial_{\mathbf r}f_1\right) = -\mathbf v\cdot \partial_{\mathbf r}\left[\ln\left(\frac{f_1}{f_{01}}\right)n_p f_1\right], \\
  \nonumber 
  && \left[1+\ln\left(\frac{f_1}{f_{01}}\right)\right]n_p \left(- u_p \partial_{x_p}f_1\right) =-\partial_{x_p}\left[\ln\left(\frac{f_1}{f_{01}}\right)n_pu_pf_1\right]\\ && \qquad +f_1\ln\left(\frac{f_1}{f_{01}}\right)\partial_{x_p}(u_pn_p),
\end{eqnarray}
and their respective integrals include
\begin{eqnarray}
  \nonumber 
  -&&\int d\mathbf v \int_0^L dx_p\int d\mathbf r_\perp\int_0^{x_p} d x\ \mathbf v\cdot \partial_{\mathbf r}\left[\ln\left(\frac{f_1}{f_{01}}\right)n_p f_1\right]\\ && =-\int d\mathbf v \int_0^L dx_p\int d\mathbf r_\perp\ v_x\left[\ln\left(\frac{f_1}{f_{01}}\right)n_p f_1\right]_{x=x_p}, \\
\nonumber    -&&\int d\mathbf r_\perp\int_0^L dx_p\int_0^{x_p} d x\ \partial_{x_p}\left[\ln\left(\frac{f_1}{f_{01}}\right)n_pu_pf_1\right]\\  && = \int d\mathbf r \left[\ln\left(\frac{f_1}{f_{01}}\right)n_pu_pf_1\right]_{x_p=x}.
\end{eqnarray}
After some manipulations, we finally arrive at 
\begin{eqnarray}
  \nonumber
  \frac{d}{dt}\int d\mathbf rdx_pd\mathbf v\ f_1n_p \ln\left(\frac{f_1}{f_{01}}\right)=&& \int d\mathbf r d\mathbf v\ (u_p-v_x)\left.\ln\left(\frac{f_1}{f_{01}}\right)n_pf_1\right|_{x_p=x}\\ &&+ \int d\mathbf rdx_pd\mathbf v\ \ln\left(\frac{f_1}{f_{01}}\right)n_pJ_{11},
\end{eqnarray}
and similarly
\begin{eqnarray}
  \nonumber
  \frac{d}{dt}\int d\mathbf rdx_pd\mathbf v\ f_2n_p \ln\left(\frac{f_2}{f_{02}}\right)=& &\int d\mathbf r d\mathbf v\ (v_x-u_p)\left.\ln\left(\frac{f_2}{f_{02}}\right)n_pf_2\right|_{x_p=x}\\ &&+ \int d\mathbf rdx_pd\mathbf v\ \ln\left(\frac{f_2}{f_{02}}\right)n_pJ_{22}.
\end{eqnarray}
It is still possible to rewrite the terms involving the piston on the last two expression in a form which will be useful soon. After some manipulations and the use of the kinetic boundary conditions \eqref{eq:bc1}-\eqref{eq:bc2}, we have 
\begin{eqnarray}
  \nonumber
    && \int d\mathbf r d\mathbf v\ (u_p-v_x)\left.\ln\left(\frac{f_1}{f_{01}}\right)n_pf_1\right|_{x_p=x}\\ && = \int d\mathbf r \int_{v_x>v_p}d\mathbf v dv_p\ (v_x-v_p)f_1f_p\left.\ln\left(\frac{f_1'}{f_{1}}\right)\right|_{x_p=x}
\end{eqnarray}
and
\begin{eqnarray}
   \nonumber
  && \int d\mathbf r d\mathbf v\ (u_x-v_p)\left.\ln\left(\frac{f_1}{f_{01}}\right)n_pf_1\right|_{x_p=x}\\ && = \int d\mathbf r \int_{v_x<v_p}d\mathbf v dv_p\ (v_p-v_x)f_2f_p\left.\ln\left(\frac{f_2'}{f_{2}}\right)\right|_{x_p=x},
\end{eqnarray}
where the primer on the functions indicates that the velocity of the respective arguments has to be replaced by its postcollision value. 

For the time derivative of the last term of Eq.~\eqref{eq:etaf}, we have
\begin{equation}
  \label{eq:79}
  \begin{split}
\frac{d}{dt}  \int & dx_p dv_p\ f_p(x_p,v_p,t)\ln\left[\frac{f_p(x_p,v_p,t)}{f_{0p}}\right]\\ &=\int dx_p dv_p\ \left(1+\ln\frac{f_p}{f_{0p}}\right)\left[-v_p\partial_{x_p}f_p+J_{p1}+J_{p2}\right] \\ & =\int dx_p dv_p\ \ln\frac{f_p}{f_{0p}}\left(J_{p1}+J_{p2}\right),
  \end{split}
\end{equation}
where we have used the conservation of particles in collisions and the fact that $f_p(0,v_p,t)=f_p(L,v_p,t)=0$. Using the expressions for the collision operators and after some manipulations, 
\begin{equation}
  \int dx_p dv_p\ \ln\frac{f_p}{f_{0p}} J_{p1}=\int d\mathbf r \int_{v_x>v_p}\left. d\mathbf v dv_p\  (v_x-v_p)f_1f_p\ln\frac{f_p'}{f_{p}}\right|_{x_p=x} 
\end{equation}
and
\begin{equation}
  \int dx_p dv_p\ \ln\frac{f_p}{f_{0p}} J_{p2} =\int d\mathbf r \int_{v_x<v_p}\left. d\mathbf v dv_p\  (v_p-v_x)f_2f_p\ln\frac{f_p'}{f_{p}}\right|_{x_p=x}.
\end{equation}

Collecting the different terms, 
\begin{equation}
  \begin{split}
    \frac{d}{dt}\mathcal H&= \int d\mathbf r \int_{v_x>v_p}d\mathbf v dv_p\ (v_x-v_p)f_1f_p\left.\ln\left(\frac{f_1'f_p'}{f_{1}f_p}\right)\right|_{x_p=x} \\ &+ \int d\mathbf r \int_{v_x<v_p}d\mathbf v dv_p\ (v_p-v_x)f_2f_p\left.\ln\left(\frac{f_2'f_p'}{f_{2}f_p}\right)\right|_{x_p=x}\\ &+ \int d\mathbf rdx_pd\mathbf v\ n_p\left[ \ln\left(\frac{f_1}{f_{01}}\right)J_{11}+ \ln\left(\frac{f_2}{f_{02}}\right)J_{22}\right].
  \end{split}
\end{equation}
As it is well known, the last term on the rhs is negative \cite{mc89}. The first term on the rhs is, using the inequality $x\ln\frac{y}{x}\le y-x$ for positive numbers $x$ and $y$, 
\begin{equation}
  \begin{split}
    &\int d\mathbf r \int_{v_x>v_p}d\mathbf v dv_p\ (v_x-v_p)f_1f_p\left.\ln\left(\frac{f_1'f_p'}{f_{1}f_p}\right)\right|_{x_p=x} \\ &\le \int d\mathbf r \int_{v_x>v_p}d\mathbf v dv_p\ (v_x-v_p)\left.\left(f_1'f_p'-f_{1}f_p\right)\right|_{x_p=x},
  \end{split}
\end{equation}
which implies, using the fact that $v_x-v_p=v_p'-v_x'$ and the kinetic boundary condition,
\begin{equation}
  \begin{split}
    &\int d\mathbf r \int_{v_x>v_p}d\mathbf v dv_p\ (v_x-v_p)f_1f_p\left.\ln\left(\frac{f_1'f_p'}{f_{1}f_p}\right)\right|_{x_p=x} \\ &\le \int d\mathbf r \int_{v_x>v_p}d\mathbf v dv_p\ (v_x-v_p)\left. f_1f_p\right|_{x_p=x} \\ & - \int d\mathbf r \int_{v_x>v_p}d\mathbf v dv_p\ (v_x-v_p)\left.f_{1}f_p\right|_{x_p=x}=0.
  \end{split}
\end{equation}
Analogously, 
\begin{equation}
  \begin{split}
    &\int d\mathbf r \int_{v_x<v_p}d\mathbf v dv_p\ (v_p-v_x)f_2f_p\left.\ln\left(\frac{f_2'f_p'}{f_{2}f_p}\right)\right|_{x_p=x} \le 0.
  \end{split}
\end{equation}

Hence, 
\begin{equation}
  \frac{d}{dt}\mathcal H\le 0.
\end{equation}

Following the argument in \cite{mc89}, for $\frac{d}{dt}\mathcal H$ to be zero, we infer the need of the distribution function to be that of the canonical distribution, Eqs. \eqref{eq:f1eq}-\eqref{eq:fpeq}.

\subsection{Proof of $\mathcal H\ge \mathcal H_0$}
Let $\mathcal H_0$ be the function $\mathcal H$ evaluated at thermal equilibrium, with Eqs. \eqref{eq:f1eq}-\eqref{eq:fpeq} and \eqref{eq:etaf}. It is 
\begin{eqnarray}
  \label{eq:difh0}
  \nonumber
  \mathcal H-\mathcal H_0= && \int d\mathbf rdx_pd\mathbf vdv_p\ f_1f_p\ln\left[\frac{f_1}{f_{1,eq}}\right]  + \int d\mathbf rdx_pd\mathbf vdv_p\ f_2f_p\ln\left[\frac{f_2}{f_{2,eq}}\right] \\ \nonumber && + \int dx_p dv_p\ f_p\ln\left[\frac{f_p}{f_{p,eq}}\right]\\ \nonumber && +\int d\mathbf rdx_pd\mathbf vdv_p\ (f_{1,eq}f_{p,eq}-f_1f_p)\ln\left[\frac{f_{1,eq}}{f_{01}}\right] \\ \nonumber && + \int d\mathbf rdx_pd\mathbf vdv_p\ (f_{2,eq}f_{p,eq}-f_2f_p)\ln\left[\frac{f_{2,eq}}{f_{02}}\right] \\ && + \int dx_p dv_p\ (f_{p,eq}-f_p)\ln\left[\frac{f_{p,eq}}{f_{0p}}\right],
\end{eqnarray}
where the suffix $eq$ denotes the equilibrium distributions. 

The first term on the rhs is
\begin{eqnarray}
\label{eq:inec1}
\nonumber
\int d\mathbf rdx_pd\mathbf vdv_p\ && f_1f_p\ln\left[\frac{f_1}{f_{1,eq}}\right]= \\ && \int d\mathbf rdx_pd\mathbf vdv_p\ f_{1,eq}f_p\left\{1+\frac{f_1}{f_{1,eq}}\left[\ln\left(\frac{f_1}{f_{1,eq}}\right)-1\right]\right\},
\end{eqnarray}
where we have used the fact that 
\begin{equation}
  \int d\mathbf rdx_pd\mathbf vdv_p\ f_{1,eq}f_p=\int d\mathbf rdx_pd\mathbf vdv_p\ f_{1}f_p.
\end{equation}
Since $1+x(\ln x-1)\ge 0$ for $x\ge 0$, the expression in Eq. \eqref{eq:inec1} is positive. A similar reasoning shows that the second and third terms of Eq. \eqref{eq:difh0} are also positive. 

The last three terms of Eq. \eqref{eq:difh0} are 
\begin{eqnarray}
  \label{eq:3t}
  -\frac{\beta}{2}&&\left[\int d\mathbf rdx_pd\mathbf vdv_p\ (f_{1,eq}f_{p,eq}-f_1f_p)m_1v_1^2\right. \\ \nonumber && + \int d\mathbf rdx_pd\mathbf vdv_p\ (f_{2,eq}f_{p,eq}-f_2f_p)m_2v_2^2\\ && + \left.\int dx_p dv_p\ (f_{p,eq}-f_p)m_pv_p^2\right],  
\end{eqnarray}
where use has been made of the form of the equilibrium distribution function and the fact that 
\begin{eqnarray}
  && \int d\mathbf rdx_pd\mathbf vdv_p\ (f_{i,eq}f_{p,eq}-f_if_p)=0, \quad i=1,2, \\ 
  && \int dx_p dv_p\ (f_{p,eq}-f_p)=0. 
\end{eqnarray}
Hence, Eq. \eqref{eq:3t} is 
\begin{equation}
  -\frac{\beta}{2}\left(E_{eq}-E\right)=0,
\end{equation}
where $E_{eq}$ is the mean energy at thermal equilibrium, with is the same as the mean energy $E$ for any other state, since collisions conserve energy. 

This way, $\mathcal H\ge \mathcal H_0$.

\section{Discussion and conclusions \label{sec:5}}
In this work, we have proposed a mesoscopic description of a system made of a cylinder including a finite number of hard spheres in $d$ dimensions divided by the adiabatic piston. We started by computing some equilibrium properties of the system as given by the canonical distribution. The analysis showed that the fluctuations of the position of the piston make the probability distribution of the gases to be spatially inhomogenous. However, a direct calculation also showed that the conditional distributions of the gases to a position of the piston are spatially homogenous, which provide a direct characterization of the global equilibrium. Furthermore, it can be shown that the conditional distributions of the gases to a position and a velocity of the piston depend on the position but not on the velocity, which reflect the presence of spatial correlation as well as the absence of velocity correlations at thermal equilibrium.

The derivation of the kinetic equations depend on few assumptions. As it is usual in Kinetic Theory, we have used the molecular chaos hypothesis by removing velocity correlations when two particles or a particle and the piston are about to collide. In addition, we have also remove some velocity correlations between particles and the piston after a collision. As already discussed, this is expected to be a good approximation if the piston suffers from many collisions in a mean free time of the surrounding particles, when the gases are dilute enough. As a result of the approximations, we ended up with a closed system of equations which solved the adiabatic piston problem, that is to say describes the evolution of the system towards thermal equilibrium where the two gases have the same pressure and all temperatures, including that of the piston, are equal. As a main difference with respect to other theories one may find in the literature, the one here is not restricted to the thermodynamic limit, nor to small particle to piston masses ratio, and include short-ranged collisions between particles.

Many of the results of the present work are easily generalized along different directions. Even though the kinetic description is for a system of hard spheres, more general interactions among particle can be considered, by replacing the collision operators $J_{11}$ and $J_{22}$ with the appropriate ones. Moreover, for most the of the collision operators conserving the number of particles, linear momentum, and energy, the $\mathcal H$-theorem still holds. More in general, we can also include dissipation, thermostats, vibrating walls, more pistons as in \cite{br12,cacesavu17}, and so on. 

An interesting aspect not addressed in this work is to precise the conditions under which the assumptions are expected to be correct. This would need a formulation including velocity correlations, which are seems to be present for small system even at thermal equilibrium \cite{cegrsavivu14}. This aspect is left for future works.

\begin{acknowledgements}
I dedicate this work to the memory of Mar\'ia Jos\'e Ruiz Montero. 
\end{acknowledgements}

\bibliographystyle{spphys}       
\bibliography{biblio.bib}   


\end{document}